\RequirePackage{fix-cm}
\RequirePackage{amsmath}

\documentclass{svjour3}

\usepackage[a4paper, left = 2cm, right = 2cm, top = 4cm, bottom = 4cm]{geometry}
\usepackage{graphicx}
\usepackage{appendix}
\usepackage{amssymb}
\usepackage{lineno}
\usepackage{array}
\usepackage{longtable}
\usepackage{natbib}
\usepackage{hyperref}
\usepackage{allrunes}
\usepackage{listings}
\usepackage{algorithm}
\usepackage{algpseudocode}
\usepackage[dvipsnames]{xcolor}
\usepackage[labelsep = period]{caption}
\usepackage[normalem]{ulem}

\makeatletter
\def\abstract{\topsep=0pt\partopsep=0pt\parsep=0pt\itemsep=0pt\relax
\trivlist\item[\hskip\labelsep
{\bfseries\abstractname}]\hskip-\labelsep}
\if@twocolumn
    
\else
\fi
\makeatother

\makeatletter
\ch@ckobl{date}{\vspace{-10pt}}
\makeatother

\def\makeheadbox{{%
\hbox to0pt{\vbox{\baselineskip=10dd\hrule\hbox
to\hsize{\vrule\kern3pt\vbox{\kern3pt
\hbox{\bfseries Journal of Quantitative Criminology}

\kern3pt}\hfil\kern3pt\vrule}\hrule}%
\hss}}}

\makeatletter
\def\makeheadbox{}

\makeatother

\smartqed 

\begin{document}

\title{Filaments of Crime: Informing Policing via Thresholded Ridge Estimation}
\titlerunning{Informing Policing via Thresholded Ridge Estimation}

\author{Ben Moews         \and
        Jaime R. Argueta, Jr.  \and 
        Antonia Gieschen
        \vspace{-35pt}
}

\institute{Ben Moews \at
              Institute for Astronomy, University of Edinburgh\\
              Royal Observatory, Edinburgh, EH9 3HJ, UK\\
              \email{b.moews@ed.ac.uk}
           \and
           Jaime R. Argueta, Jr. \at
              School of Criminal Justice, University of Cincinnati\\
              2600 Clifton Ave, Cincinnati, OH 45221, USA\\
              \email{arguetjr@mail.uc.edu}
           \and
           Antonia Gieschen \at
              Business School, 
              University of Edinburgh\\
              29 Buccleuch Place, Edinburgh, EH8 9JS, UK\\
              \email{antonia.gieschen@ed.ac.uk}
}

\maketitle

\vspace{-50pt}
\begin{abstract}

In this study, we investigate the potential for optimizing hot spot patrol routes through density ridge estimation. We explore the application of an extended version of the subspace-constrained mean shift algorithm by using 2018 and 2019 Part I crime data from Chicago. Ultimately, the goal of mapping hot spots is to show concentrations of crime, thus targeting the epicenters only focuses on one problem area. For this reason, we refine patrol optimization to focus on the critical ridges in hot spots. In doing so, we extract density ridges of 2018 to early 2019 Part I crime incidents from Chicago to demonstrate that nonlinear mode-following ridges agree with broader kernel density estimations. We create multi-run confidence intervals and show that our patrol templates cover around 94\% of incidents for 0.1-mile envelopes around ridges, and deliver evidence that ridges following crime densities enhances the efficiency of patrols. Our post-hoc tests show the stability of ridges, thus offering an alternative patrol route option that is effective and efficient.

\smallskip



\vspace{20pt}
\end{abstract}

\section{Introduction}
\label{intro}

Investigations of hot spot policing tactics find that focused efforts on problem areas, such as staying at a block, effectively reduce crime \cite{braga2014effects, corsaro2019implementing}.  Related research finds that 15 minutes of police presence in a given hot spot significantly decrease both calls for service and Part I crimes \cite{koper1995just, telep2014much}. These practices suggest that the stability of crime places allows for the optimization of tactics. Specifically, one way to optimize patrol routes is to a focus on the spatial aspects of hot spots, for example targeting streets and assigning prevention resources to them \cite{Camacho-Collados2015}.

We explore the optimization of hot spot patrols by identifying density segments as targets for crime prevention. Hot spots have high-density centers depending on the parameters defined by the analyst \cite[pp. 356–357 ff. of][]{eck2012place}. Practitioners and police officers rely on spatial analytics to identify hot spots that dictate their patrols, often over-emphasizing the epicenter's value. This emphasis potentially over-patrols the core area and under-patrols the surrounding areas \cite{Eck2005}. Therefore, we propose and investigate an optimized patrol algorithm that identifies crime ridge densities in the surrounding hot spot to allow for a spread of patrol.

Previous scholars explored patrol route optimization through a variety of techniques, including multi-agent-based simulations \cite{Fukunaga1975}, machine learning \cite{li2011police, marchant2018cox}, and graph theory and evolutionary computing \cite{chawathe2007organizing, al2016automatic}. These studies consistently  show that route optimization is a feasible task, and account for resources and time. The primary issue among these methods is their limited application, as more complex approaches do not equate to  effectiveness. Covering each street and hot spot by spending little time between places may lead to hot spots not meeting the required dosage or frequency of visits \cite{kringen2017assessing}. In turn, these patrols may have limited effectiveness or backfire \cite{linning2018weak}.

Both practices, hot spot patrols and patrol optimization, tend to focus on the hot spot's epicenter. Thus, the mismatch between the two bodies creates a gap in efficiency and effectiveness. This mismatch demonstrates three problems. First, patrol algorithms' implementations fixate on a single spot for hot spot patrols \cite{Eck2005}. Secondly, common practices of identifying hot spots lack patrol direction \cite{chainey2008utility, ratcliffe2010crime}. Thirdly, patrol optimization algorithms propose a comprehensive list of all routes to be covered, thus under-patrolling areas.

In this paper, we suggest a way to bridge this gap. Recently, advances in statistics to perform density ridge estimation have enabled the construction of ridges that follow high-density areas, or modes, of a distribution and allow for higher-dimensional extensions \cite{Ozertem2011, Chen2015c}. In effect, this means the extraction of curvilinear structures, or `filaments', that show high-density pathways reflecting an underlying distribution. As such, density ridges are different from mode-finding hot spot approaches, offering the identification of a connected network while identifying finer-grained structures less prone to oversmoothing risks \cite{Genovese2014}.

For this reason, the present study is exploratory. We seek to address the previous shortcomings of patrol optimization by applying methods from neighboring disciplines. We focus on and extend the subspace-constrained mean shift (SCMS) algorithm \cite{Ozertem2011} in order to introduce the concept of density ridges to the field of criminology. The identification of ridges will allow law enforcement to efficiently patrol routes in hot spots and surrounding areas. Thus, we contribute to the greater literature by uniting patrol optimization work and ridge estimation in hot spots to select patrol routes. The introduction of density ridges demonstrates advantages by offering efficiency as well as more equitable and focused patrols through the inclusion of finer-grained information on the density landscape. We show that these density ridges cover more problem segments in hot spots compared to hot spot policing or placing police personnel at single locations, and thus are an effective tool more suitable for crime prevention patrol.

We make use of Chicago Part I crime incident data from 2018 to develop and illustrate the application of ridge estimation through computational experiments. Additionally, we use data from January to May 2019 to test for predictive accuracy in coverage, as well as for convergence consistency, with multi-run confidence intervals and additional experiments for alternative method comparisons. Chicago offers an ideal data set that allows for mapping and testing, has a typical urban street network, and provides plentiful crime data. Thus, this paper's empirical work assesses the potential of patrol optimization in urban cities while going beyond current good policing practices.

\section{Literature review}
\label{sec:lit_review}

\subsection{Hot spots}
\label{sec:policing}

Over the past two decades, scholars have confirmed that large numbers of calls for service concentrate within 3–5\% of a given city \cite{sherman1989hot, sherman1995general, braga2014effects}. Related research finds that hot spots chronically persist for longer than a decade in 5\% of block-long street segments \cite{weisburd2004trajectories}. Since then, the `Law of Crime Concentration' was coined \cite{weisburd2015law}, which refers to the concept where crime concentrates in specific small areas of any city or year. Later works find that hot spots vary in size for different types of crimes, for example gun-related crime  \cite{braga2010concentration}, robberies \cite{braga2012hot}, and other major crimes \cite{haberman2017overlapping}.

Findings from these studies enable researchers and practitioners in two ways. The first is testing techniques on stable hot spots and investigating which policing strategies can be the most effective.  Patrolling hot spots is reported to not disperse crime to neighboring geographic locations \cite{braga2007policing}. Instead, deterrent effects are diffused to nearby streets, making efforts in patrolling hot spots a successful endeavor \cite{braga2014effects}. Recently, research raises issues with the amount of patrolling in hot spots, pointing out a possible hermetic effect \cite{linning2018weak}. The latter work suggests that if a hot spot does not meet a specific dosage of patrol presence, there may be an increase in crime. Thus, under-policing or even over-policing areas can backfire and result in increases of criminal activity. Similarly, scholars argue that patrols should focus on fewer visits of longer duration at hot spots rather than hitting them randomly and often \cite{williams2017frequency}.

The research mentioned above developed in parallel to methods for estimating hot spots. With about 75\% of agencies using the hot spots policing approach, most use kernel density estimations to identify intervention areas \cite{national2018proactive, mastrofski2015police}. While kernel methods show great success, they only focus on singular cells and spaces instead of the surrounding problem areas or opportunities. This lead to the suggestion that there should perhaps be more to just plotting densities of crime areas \cite{eck1997those}. To address this, a risk terrain model has been presented \cite{caplan2011risk}, applying forecasting of opportunity structures throughout a geographic space and building on prior kernel density work. This does, however, still include possible pitfalls of an over-reliance on the epicenter of cells to suggest patrol work. Hence, the risk terrain model and kernel density models still apply a static approach for placing police on dots, or epicenters, with little consideration for the full spatial range identified.

In summary, hot spot policing provides a way for police departments to reduce crimes by patrolling problem areas effectively, but there are limitations to the use of kernel methods. Hot spots illustrate varying levels of crime concentration over a geographic landscape. The empirical work described in this paper suggests that the decision of where the line is drawn by analysts to define a hot spot may vary the amount of attention. Thus, hot spot patrols may benefit from a defined route that exhibits optimal routes to target crime prevention resources so they do not focus on just one area. 

\subsection{Patrol optimization}
\label{sec:patrolling}

For strategic planning, law enforcement makes use of hot spots to identify problem areas to patrol.The visible presence of patrols in a community is one of the key components in reducing crime, especially in hot spots. For this reason, the identification of routes in hot spots is relevant due to patrols being constrained by street networks \cite{menton2008bicycle}. Furthermore, given the scarcity of police resources, the efficient allocation of proactive patrols is crucial, and an optimal dosage of police presence at these hot spots needs to be applied. Police agencies identify hot spots with spatial ellipses, grid mapping, thematic mapping, kernel density methods, Getis-Ord Gi*, and point processes for spatial analytics \cite{chainey2008utility, ratcliffe2010crime, Xue2006}. That being said, advanced route planning based on proper hot spot estimations still lags behind most current research.

Scholars have recently turn to algorithms. Patrol optimization deals with identifying optimal routes so that officers target hot spots efficiently. While these methods are complex, they offer the potential to dynamically shape patrol routes to service each call, problem area, or assignment. For example, recent work using the ant colony optimization algorithm and Bayesian methods \cite{chen2015designing, Furtado2009} shows the practical utility of efficiently hitting each hot spot in an optimal manner. In another approach, dynamic modeling is used by assuming that offenders will predict patrol routes \cite{paruchuri2008playing}, demonstrating the model's ability to determine optimal paths that balance predictable and unpredictable street network paths. Related research suggests the potential to decrease criminal activity and the public's fear of crime by modeling patrol routes illustrating the shortest Hamiltonian cycle for visiting each location in a city \cite{chevaleyre2004theoretical2}.

Additional facets of patrol optimization consider limited patrol resources and take on a variety of approaches. These include the application of patrol optimization using a cost-benefit analysis, maximizing the coverage of hot spots and accounting for the paths between streets and places \cite{chawathe2007organizing}, as well as a multi-agent-based algorithm to design efficient patrol strategies \cite{reis2006gapatrol}. The latter simulation models a city's road network to find optimal routes to minimize crime in a city. Further work studies changing offenders' opportunity structure \cite{furtado2006using}, while related efforts simulate changing problem places that adapt to patrol routes \cite{melo2005analyzing}. Finally, similar applications look at how district models can be optimized to adequately distribute calls for service or incidents in a given jurisdictions  \cite{liberatore2020police, mitchell1972optimal, bodily1978police, piyadasun2017rationalizing}. The shortcoming of each of these works is that they focus on cost efficiency and formulation of routes. Only few works include the importance of hitting potential problem places, although these studies do not account for the quality of patrols \cite{reis2006gapatrol, melo2005analyzing}.

Even with progress underway, there are still several limitations that patrol optimization studies fail to consider in their analyses. To our knowledge, none of the existing patrol optimization articles and hot spot research meet quality patrol needs while being efficient. Additionally, shortening the scope of patrol optimization to deal with one problem appears to be a feasible approach to managing proper efficiency in crime prevention at hot spots. The modeling of optimal patrol routes and simulated agents to combat problem places and limited resources is still in the early stages. While valuable for researching the impacts of policing strategies, real-world applications of optimal patrol routes are, thus, severely limited.

\section{Data and methods}
\label{sec:data_and_methods}
\subsection{Crime incident data}
\label{sec:data}

We use the Chicago Data Portal\footnote{\url{https://data.cityofchicago.org/}}, an open-access data service. The portal features a complete dataset of reported crime incidents from 2001 to the present day, covering over 17 years, with the exception of murders where data exist for each victim. The crime incident data are provided by the Citizen Law Enforcement Analysis and Reporting (CLEAR) system of the Chicago Police Department. CLEAR's choice offered an ideal data set that allowed for mapping, testing, and near-current crime events.

After obtaining the dataset, we extract all entries pertaining to 2018, and retain only three variables of interest; the primary crime type and the coordinates of the reported crime's location. We plot the coordinates using ArcMap 10 and TIGER street centerlines projected to Geographic Coordinate System North American 1983 as spatial reference data. After this step, we omit all entries for which at least one of the retained variables is not present. This omission for missing data leads to the data for 2018 being reduced from 178,659 to 177,669 entries, resulting in a negligible loss of around 0.5\% of data points.

In this work, we focus on Part I offenses as defined by the Uniform Crime Reports\footnote{\url{https://www.ucrdatatool.gov/}} (UCR). Our choice of Part I crimes reflects both the high priority placed on this type of crime and its reliability, as well as its prior use in patrol route optimization studies \cite{barnett2014nation, chen2015designing, chen2017developing}. In this context, aggravated assault, forcible rape, criminal homicide, and robbery are Part I violent crimes, whereas arson, burglary, larceny-theft, and motor vehicle theft are Part I property crimes. We extract these eight primary types from the preprocessed dataset, which leaves us with 78,894 incidents of Part I crimes in Chicago during the year 2018, an overview of which is provided in Tab.~\ref{tab:dataset}. In order to keep our algorithm's runtime low, and given that we are interested in keeping the overall density profile of crime incidents, we use uniformly-random sampling to reduce the dataset to 5,000 data points and show, in Section~\ref{sec:results}, the sufficiency of the sample in a predictive case.

\begin{table}
\begin{center}
\caption{Part I crime incident numbers for Chicago during the year 2018. Different primary crime types are listed separately, with entries descending by the number of reported incidents.}
\label{tab:dataset}
\begin{tabular}{ll}
\hline\noalign{\smallskip}
Primary crime type & Number of data points \\
\noalign{\smallskip}\hline\noalign{\smallskip}
Larceny-theft & $42,423$ \\
Aggravated assault & $13,843$ \\
Burglary & $7,821$ \\
Motor vehicle theft & $6,641$ \\
Robbery & $6,525$ \\
Forcible rape & $1,013$ \\
Criminal homicide & $386$ \\
Arson & $242$ \\
\noalign{\smallskip}\hline
\end{tabular}
\end{center}
\end{table}

\subsection{Subspace-constrained mean shift}
\label{sec:scms}

Our approach is an extension of the subspace-constrained mean shift algorithm (SCMS), a density ridge estimation method that has been further extended in application areas described below. Following the examples of other scholars, we further adapt and extend the algorithm for a criminological context. The SCMS algorithm can be applied to crime patterns to extract route templates from high-density areas.

In order to provide readers with the background of the employed method in more depth, a short overview of the mathematical foundations is required. Given a probability density function $p : \mathbb{R}^d \rightarrow \mathbb{R}$ of dimensionality $d$, as well as a corresponding gradient $\nabla p(x)$ and a Hessian $H(x)$, let $v = \{ v_1, v_2, \dots, v_d \}$ be the eigenvectors of $H(x)$ corresponding to eigenvalues $\lambda = \{ \lambda_1, \lambda_2, \dots, \lambda_d \}$ sorted in descending order. Defining $\Lambda(x)$ as the diagonal matrix with $\lambda$ along the diagonal, and with the eigendecompostion $H(x) = U(x) \Lambda(x) U(x)^{\top}$, we let $v'$ be the columns of $U(x)$ associated with the $d - 1$ smallest entries in $\lambda$. In addition, let $L(x) \propto L(H(x)) = v' v'^{\top}$ be a projection on the linear space of the columns in $v'$, then the projected gradient is defined as $\nabla p(x) = L(x)g(x)$. For a map $\xi : \mathbb{R} \rightarrow \mathbb{R}^d$, the ridge $R$ can be expressed as $R = \{ x : ||G(x)|| = 0, \lambda_{d + 1} (x) < 0 \}$ \cite{Ozertem2011, Genovese2014}. In other words, a density ridge is a local density maximization in the normal direction given by the Hessian. While the above provides a bare-bones definition, we refer the interested reader to \cite{Genovese2014} for a more detailed introduction to non-parametric ridge estimation.

Kernel density estimation, which is also known as the Parzen-Rosenblatt window, is a non-parametric statistical method to estimate probability density functions \cite{Rosenblatt1956, Parzen1962}. The most common choice, and the one used in our approach, is the radial basis function (RBF) kernel, also known as the Gaussian kernel, with $\mathcal{K} (x) = (1 / \sqrt{2 \pi}) \exp(-0.5 x^2)$. The SCMS algorithm \cite{Ozertem2011} is a KDE-based non-parametric iterative approach to estimate the ridges of a probability density function in the context of self-consistent smooth curves using $\nabla p(x)$ and $H(x)$. While the literature on applications since its recent introduction is sparse, the algorithm has been applied to neuroscience \cite{Bas2011} and road networks \cite{Miao2014}, as well as in astronomy~\cite{Chen2015a, Chen2015b, Chen2015c, Chen2016, Chen2017, He2017, Hendel2019, Moews2020}. Specifically, the method is extended with thresholding \cite{Chen2015a} for the application to cosmic web reconstruction, using a KDE over the dataset to counteract the effect of areas with low probability densities.

\begin{algorithm}
\caption{SCMS with thresholding}
\begin{algorithmic}[1]
\State \textbf{Input:} Coordinates $\theta$, bandwidth $\beta$, threshold $\tau$, iterations $N$
\State \textbf{Output:} Density ridge point coordinates $\psi$
\Procedure {SCMS}{$\theta$, $\beta$, $\tau$, $N$}
\State $\kappa(x) \longleftarrow \mathrm{KDE}_{\mathrm{RBF}} (\theta, \beta)$
\State $\mathcal{\psi} \longleftarrow \mathcal{\psi} \sim U((\min(\theta_{\ast, 1}), \max(\theta_{\ast, 1})), (\min(\theta_{\ast, 2}), \max(\theta_{\ast, 2})))_{|\theta|}$
\State $\psi \longleftarrow \forall y \in \psi : \kappa(y) > \tau$
\For {$n \longleftarrow 1, 2, \dots, N$}
\For {$i \longleftarrow 1, 2, \dots, |\psi|$}
\For {$j \longleftarrow 1, 2, \dots, |\theta|$}
\State $\mu_j = \frac{\psi_i - \theta_j}{\beta^2}$
\State $\sigma_j = \mathcal{K}_{\mathrm{RBF}} \left( \frac{\psi_i - \theta_j}{\beta} \right)$
\EndFor
\State $H(x) = \frac{1}{|\theta|} \sum_{j = 1}^{|\theta|} \sigma_j \left( \mu_j \mu_j^{\top} - \frac{1}{\beta^2} \mathbb{I} \right)$
\State $v, \lambda \longleftarrow v, \lambda \mathrm{ \ from \ eigendecomposition \ eig} (H(x))$
\State $v' \longleftarrow \mathrm{entries \ in \ } v \mathrm{ \ corresponding \ to \ sort}_{\mathrm{asc}} (\lambda)_{1, 2, \dots, d - 1}$
\State $\psi_i \longleftarrow v' v'^{\top} \frac{\sum_{j = 1}^{|\psi|} \sigma_j \theta_j}{\sum_{j = 1}^{|\psi|} \sigma_j}$
\EndFor
\EndFor
\State \textbf{return} $\psi$
\EndProcedure
\end{algorithmic}
\label{alg:scms}
\end{algorithm}

The convergence properties of the SCMS algorithm have been analyzed \cite{Ghassabeh2013}, showing that the method inherits some properties of the previous mean shift algorithm \cite{Fukunaga1975}, most importantly its monotonicity and the convergence of density estimates along the output sequence, together with other properties that offer theoretical guarantees for stopping criteria. For an up-to-date contextualization of the approach in the broader field of topological data analysis, we refer the reader to suitable overview \cite{Wassermann2018}, as well as to a more general analysis of non-parametric density ridge estimation \cite{Qiao2016}. In addition, a study of, as well as extensive proofs for, ridge estimation from a geometrical perspective have been conducted \cite{Genovese2012}.

\subsection{Modifications and extensions}
\label{sec:modifications}

In addition to providing a fast pure-Python implementation of the SCMS algorithm, with thresholding implemented in line 6 of Alg.~\ref{alg:scms}, we introduce multiple modifications of the methodology tailored to geospatial data and applications in criminology.

An optimal bandwidth calculation for crime incident data has been introduced earlier \cite{Williamson1999}, based on the average distance of each coordinate to its nearest $k$ neighbors, averaged over all coordinates in the dataset \cite{Eck2005}. For the distance $\textarc{d} (\theta_i, \theta_j)$ between two coordinates of a dataset $\theta$, and with the number of nearest neighbors $k$, the calculation of the optimal bandwidth $\hat{\beta}$ takes the form of the following equation: 
\begin{equation}
\label{eq:bandwidth}
\hat{\beta} = \frac{1}{k |\theta|} \sum\limits_{i = 1}^{|\theta|} \sum\limits_{j = 1}^{k} \textarc{d} (\theta_i, \theta_j)
\end{equation}
This approach is related to the \textit{k}-nearest neighbors (\textit{k}-NN) algorithm, a non-parametric statistical method commonly applied to regression and classification problems \cite{Cover1967}. We make use of this calculation to provide an optimized default bandwidth for our method. Without a bandwidth optimization, the bandwidth would need to be set manually by the user, which would lead to problems in both directions. Either the coverage would be diminished due to a too large bandwidth, resulting in ridges that follow an overly broad density profile, or the ridges would present a too fine-grained net of substructures that would mathematically provide good coverage, but not be practical for patrolling.

While the Euclidean distance is a staple in geometric calculations, its use can lead to distorted measures when applied to geospatial coordinates over sufficiently large distances. In this context, the orthodromic distance is the shortest path between two coordinates on a sphere, measured along the sphere's surface. As such, it provides a sufficiently realistic way to calculate distances as geodesics on an approximated shape of the Earth. While police patrolling is, in practice, a regional problem and local topology outweighs the curvature of our planet, a negligible difference in computational costs allows the resulting software to be applicable to other, more large-scale challenges in other fields.

The haversine function of an angle $\alpha$ is a numerically better-conditioned for small geodesic distances than using the spherical law of cosines. The haversine formula \cite{Inman1835} makes use of that function and provides a way to calculate the orthodromic distance suitable for our purposes in that it remains accurate on small-scale local distances and stays applicable on larger scales. Denoting the latitudes and longitudes separately, the haversine distance between points $\theta_1$ and $\theta_2$ is then:
\begin{equation}
\label{eq:haversine_distance}
\textarc{d}_{\mathrm{hav}} (\theta_1, \theta_2) = \mathrm{hav} (\theta_{2, 1} - \theta_{1, 1} + \cos \theta_{1, 1} \cos \theta_{2, 1} \mathrm{hav} (\theta_{2, 2} - \theta_{1, 2}))
\end{equation}
One interesting point to note is that the applicability of the haversine distance directly translates to projected astronomical observations, although with flipped horizontal axes, as the sky in the latter is viewed as a sphere with the Earth at its center. This is done in an application of our implementation \cite{Moews2020}, making direct use of our work across fields as a result, and showing the interdisciplinary applicability of methodological and software developments between fields. In addition to its use for the SCMS algorithm's iterative updates, we also use this distance for the \textit{k}-NN approach of calculating an optimal bandwidth, replacing $\textarc{d} (\theta_i, \theta_j)$ in Eq.~\ref{eq:bandwidth} with $\textarc{d}_{\mathrm{hav}} (\theta_i, \theta_j)$ from Eq.~\ref{eq:haversine_distance}.

Well-approximated density ridges require the SCMS algorithm to run over a sufficient number of iterations. Since a trial-and-error approach is not the most time-efficient way of using the algorithm, we implement a convergence check that uses the mean shift update in Alg.~\ref{alg:scms}. Let the update be denoted as $\phi_{n, i}$, for iteration $n$ and $j \in \{ 1, 2, \dots, |\psi| \}$ for ridge candidate points $\psi$, then the calculation takes the following form:
\begin{equation}
\label{update}
\phi_{n, i} = v' v'^{\top} \frac{\sum_{j = 1}^{|\psi|} \sigma_j \theta_j}{\sum_{j = 1}^{|\psi|} \sigma_j} - \psi_i
\end{equation}
We then introduce the convergence criterion, for a convergence threshold $c$, as the absolute difference between an iteration's current update and the last iteration's update not exceeding the convergence threshold, meaning that $||\phi_{n - 1, i} - \phi_{n, i}|| \leq c$. Without this introduced convergence criterion, the number of iterations would, as in the original SCMS algorithm, need to be set manually by the user. This poses the challenge of correctly guessing the number of required iterations to create well-defined ridges, as too small a value would result in fuzzy `clouds' along the ridges, as opposed to ridge lines. The value would thus need to be set rather large in order to avoid this issue, hoping to overshoot the necessary but unknown value, which would, even if successful, increase the computational costs and thus the runtime of the algorithm.

Lastly, practitioners in criminology are often primarily interested in hot spots, focusing their efforts on regions with high probability densities. In order to enable this use of our method, we propose a cut-off functionality to return only ridge estimates in regions with a high number of data points in comparison to the dataset. For a given percentage value $p$, the KDE in the SCMS algorithm is used to only retain ridge estimate points above the $(100 - p)^\mathrm{th}$ percentile of the dataset's estimated probability density function. This means that the ridge estimate points $\psi$ are, for a bandwidth $\beta$ and a Gaussian-kernel KDE, reduced to a subset $\psi'$:
\begin{equation}
\begin{split}
\label{eq:percentage}
\psi' &= \hat{\psi} \in \psi : \mathrm{KDE}_{\mathrm{RBF}}(\hat{\psi}, \beta) \geq \gamma, \\
 &\mathrm{with \ } \gamma = \min \left( \mathrm{sort}_{\mathrm{desc}} \left( \mathrm{KDE}_{\mathrm{RBF}}(\psi, \beta) \right)_{1, 2, \dots, \lfloor \frac{p}{100} |\psi| \rfloor} \right)
\end{split}
\end{equation}
This approach allows for the exclusive retention of ridge estimates that fall within regions of high probability densities, effectively slicing the density landscape horizontally at the required percentage level and extracting the ridge estimate points that can be found on the remaining landscape. The advantage of this extension is that a top-percentage level of crime density can be freely chosen to concentrate hot spot policing efforts on a highest-density subset of areas in line with additional considerations by the respective practitioners.

We introduce a pure-Python software tool for \textit{density ridge estimation describing geospatial evidence} (DREDGE), written for Python 3. The tool itself is available on, and can be installed via, the Python Package Index (PyPI).\footnote{\url{https://pypi.org/}} We also provide the complete code for DREDGE in a public repository\footnote{\url{https://github.com/moews/dredge}}, accompanied by documentation, a quickstart tutorial, and a use case featuring example code.

\section{Results}
\label{sec:results}

\subsection{Primary experiment and visualization}

\begin{figure}[!htb]
    \centering
    \includegraphics[width=\textwidth]{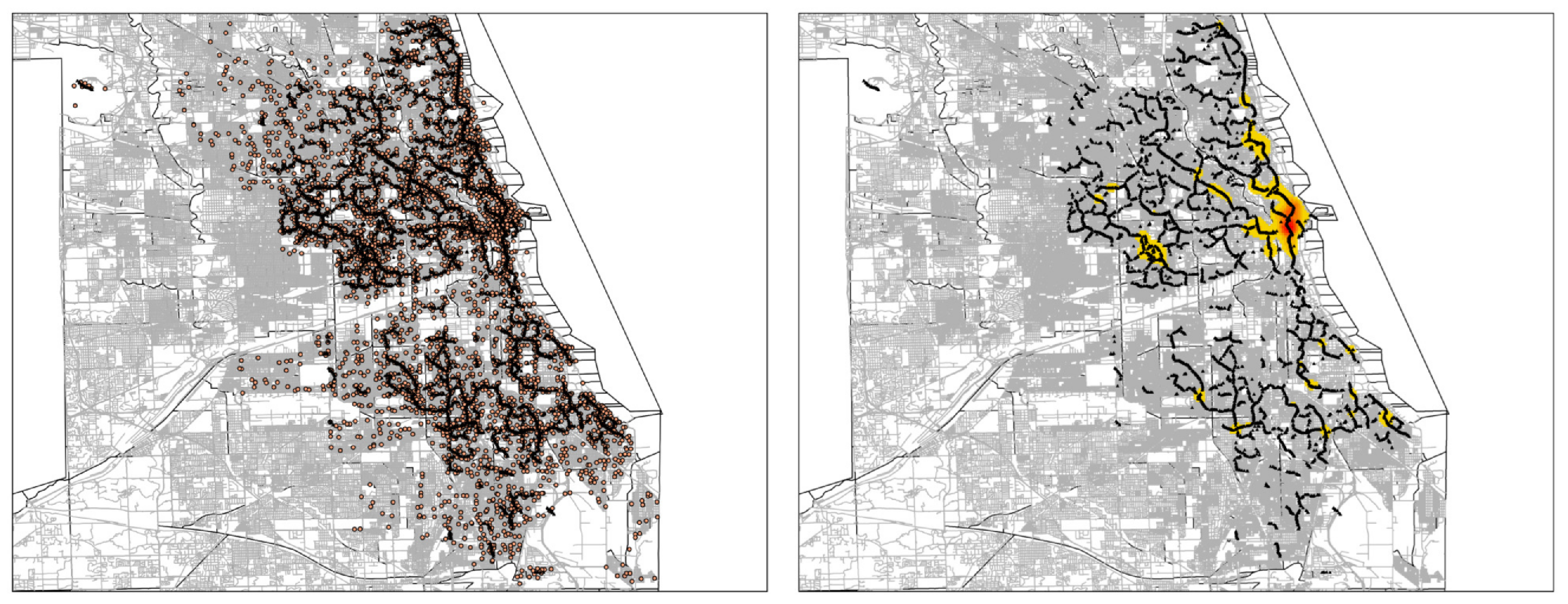}
    \caption{Full density ridges extracted from reported Part I crime incidents for the City of Chicago during 2018. The left panel adds a sample of underlying coordinates, whereas the right panel adds a kernel density estimation (KDE) for the samples in the left panel.}
    \label{fig:full_combined}
\end{figure}

The theoretical work on hot spots and direct patrols has been widely applied and studied within the field of criminology \cite{braga2014effects}. Current applications of patrol routes include patrols that are mainly planned using street network models and KDE \cite{mamalian1999use, ratcliffe2004crime}. Our work seeks to capitalize on that aspect through density ridge estimation. The density ridges obtained through this experiment with Chicago's 2018 Part I crime incidents are shown in both panels of Fig.~\ref{fig:full_combined}. In the left panel, we additionally show a sample of 5,000 coordinates of reported crime incidents, the same size as used by the DREDGE run. In the right panel, we overlay the density ridges with a KDE based on the same optimal bandwidth used by our method, demonstrating the center-line compliance of ridges with hot spots identified by traditional approaches. We show how DREDGE results line up with the underlying data as well as KDE outputs to demonstrate how our results follow high-density areas identified with this alternative approach, using the latter as a comparison baseline.

The implementation of our method described in this paper is run with default values, allowing the software to make use of its adaptive behavior. We run this experiment on an Intel Core i7-5600U CPU with 2.60 Ghz, two cores, and four threads, on a machine featuring a sufficient 8 GB of RAM and resulting in a runtime of 6 minutes and 26 seconds. The algorithm was not parallelized, running in a single-threaded fashion to gauge the out-of-the-box performance, although low-level parallelization on a CPU can be easily implemented with the `multiprocess' package.\footnote{\href{https://pypi.org/project/multiprocess/}{pypi.org/project/multiprocess}}

 \begin{figure}[!htb]
    \centering
    \includegraphics[width=\textwidth]{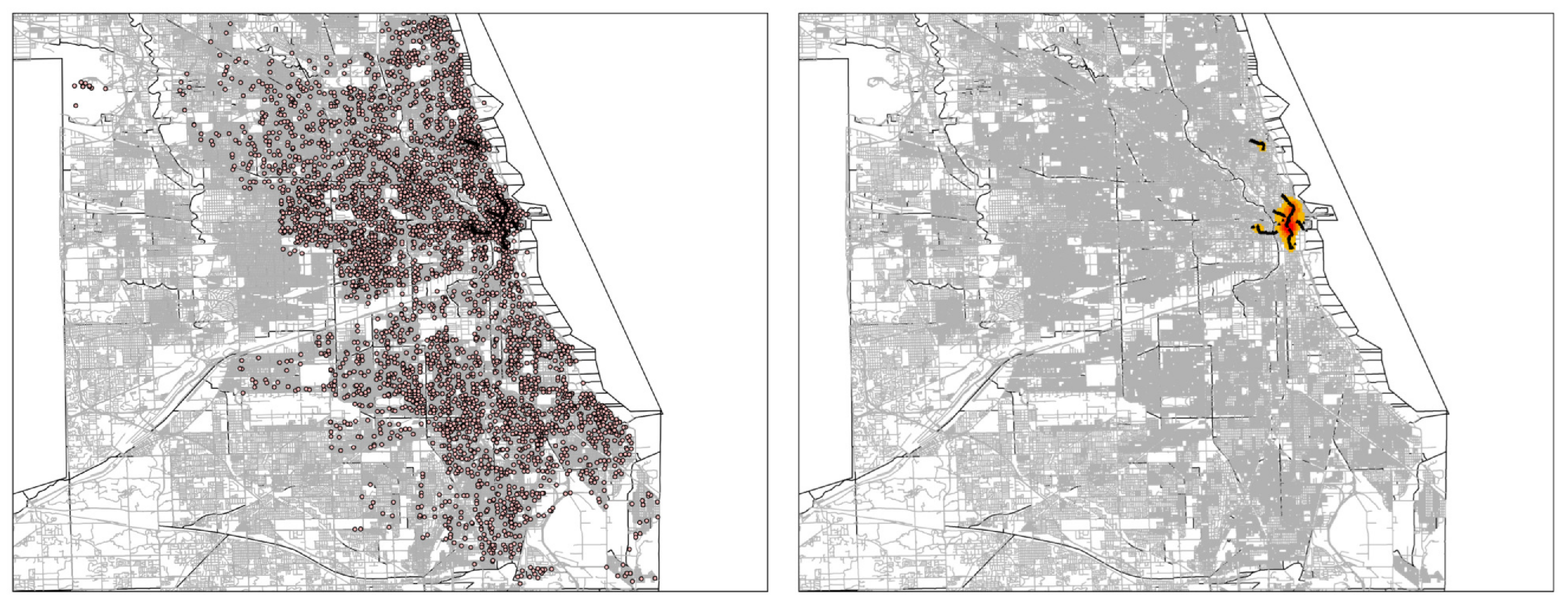}
    \caption{Partial density ridges extracted from reported Part I crime incidents for the City of Chicago during 2018. The left panel adds a sample of underlying coordinates, whereas the right panel adds a kernel density estimation (KDE) for the ridge-related hot spots.}
    \label{fig:part_combined}
\end{figure}

The practical application and ease of policing hot spots has allowed police departments to patrol specific areas more readily \cite{weisburd2005diffusion}. Capitalizing on hot spots, we make use of DREDGE's ability to retrieve density ridges from a specified level of high-density areas, as discussed in Section~\ref{sec:modifications}. Fig.~\ref{fig:part_combined} shows the respective top-percentage ridges retrieved through this experiment. Both panels show partial density ridges, making use of the built-in threshold functionality set to 5\% for density ridges covering the region above the 95$^\mathrm{th}$ percentile of the incident density distribution. As in Fig.~\ref{fig:full_combined}, the left panel additionally shows a sample of 5,000 coordinates of reported crime incidents, the same size as used by the DREDGE run, to show the relation of ridges to the underlying dataset, whereas the right panel overlays the density ridges with a KDE estimate for the data points relevant to the top 5\% ridges as a comparison baseline.
 
Due to the same underlying analysis, the high-density area highlighted through the KDE visualization in Fig.~\ref{fig:full_combined} corresponds to the hot spot singled out in Fig.~\ref{fig:part_combined}. This location falls within the Near North Side and Loop areas of downtown Chicago, known as tourist and shopping destinations. An obvious interpretation of this high-density accumulation of data points relies on the considerable overrepresentation of larceny-theft in our dataset, as these areas provide ample opportunity for such crimes, combined with scaling effects due to the number of people frequenting them, which we confirmed for over a fifth of the larceny-theft reports occurring in these areas.

\subsection{Post-hoc analyses and comparisons}

To test the predictive accuracy and stability of density ridges over time, we extract another dataset from the Chicago Data Portal. The procedure remains the same as in Section~\ref{sec:data}, but with data for the year 2019 until the end of May. This amounts to 38,205 preprocessed Part I crimes. For this experiment, we use the complete dataset without subsampling, in addition to confidence intervals for multiple runs, to assess the stability of the algorithm's performance. We employ the ridges extracted from the previous 2018 data to simulate route optimization. We measure the distance to the nearest density ridge and calculate the percentage of incidents falling into this envelope around ridges. In doing so, we quantify the amount of incidents in 2019 that happen near a route template based on 2018 data.

This approach is related to a hit rate, or the percentage of crime incidents occurring within an area of a certain size \cite{Chainey2008}. Related work \cite{Bowers2004} proposes a search efficiency rate that counts the number of events per square kilometres, although this approach lacks comparability between different study areas. Our choice to measure the percentage of overall crime incidents within envelopes around ridges bears the closest resemblance to the prediction accuracy index (PAI) \cite{Bowers2004}. The PAI computes the percentage of crime incidents within a predicted area, divided by the percentual size of the predicted areas in relation to the respective study area. Notably, accounting for the predicted area size is not a concern in our ridge-specific approach, which predicts curvilinear filaments instead of areas. Instead, our measurement's equivalent is the envelope width around ridges, which requires the calculation of the crime incident coverage for varying widths in order to accurately represent the ridges' success.

\begin{figure}[!htb]
    \centering
    \includegraphics[width=\textwidth]{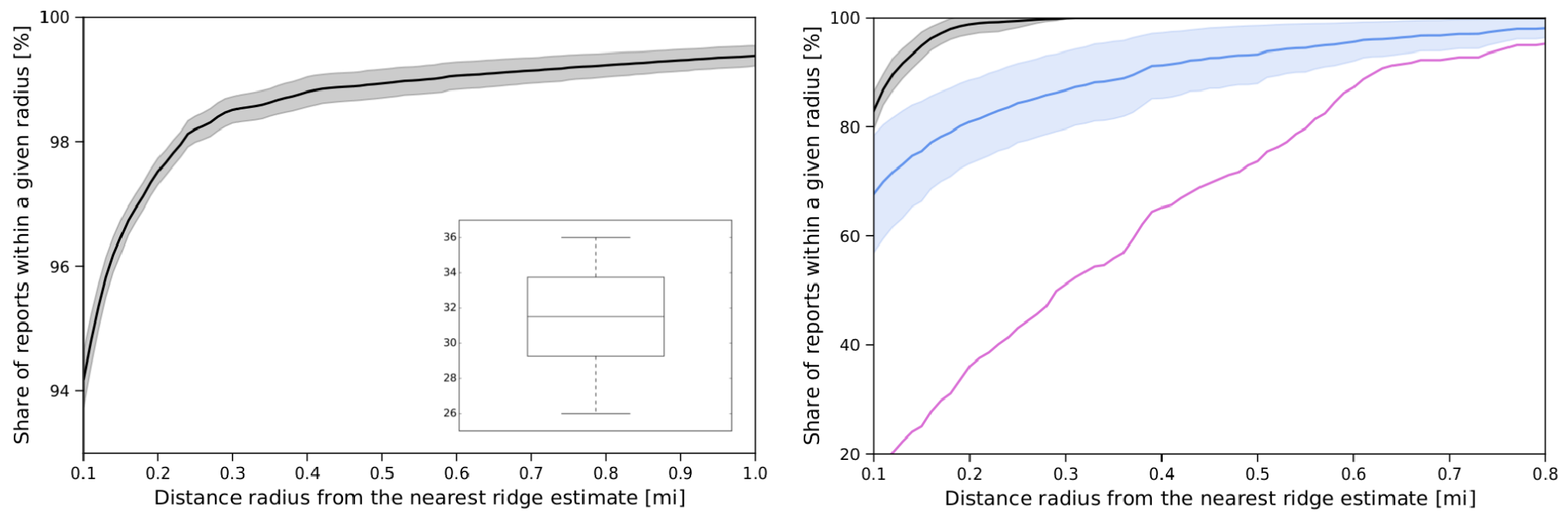}
    \caption{\textit{Left:} Distance-based coverage for Part I crimes in January to May 2019 in the City of Chicago, for ridges calculated with data from 2018 and 95\% confidence intervals in lighter shades. The subplot in the lower right corner shows a box-and-whiskers plot for the number of iterations to convergence. \textit{Right:} The same data as on the left, but results are shown for clustered hot spots in top-5\% areas, with ridges placed within them in black, random patrols within the hot spots and including the hot spot center in blue, and hot spot centers exclusively in purple.}
    \label{fig:future_experiment}
\end{figure}

We compute this experiment for distances in the interval $[0.1, 1]$ in miles, in steps of 0.01 miles, and repeat each experiment for each distance for a total of 10 times to recover confidence intervals. Each of these 10 runs per distance step is based on a random subsample of reports from the year 2018, with a different random seed each time, to validate the efficacy of a comparatively small subsample of 5,000 data points. In addition, we measure the number of iterations required each time to test the suitability of the convergence criterion introduced in Section~\ref{sec:modifications}. We also measure the same distance envelope coverage for the top 5\% of crime report densities as shown in Fig.~\ref{fig:part_combined}, and with the same ridges as depicted there, for which we identify five clusters via the mean shift algorithm and use 5\%-thresholded crime report coordinates. Using random within-cluster connections with interpolation, we create random routes per identified hot spot to compare the predictive coverage of our approach to random patrols within hot spots, as well as with the use of solely the hot spot center as a point of reference.

Fig.~\ref{fig:future_experiment} shows the results of these experiments. The black line in the left-hand plot depicts the share of Part I crime reports in the City of Chicago from 2019 data on the vertical axis, depending on the size of the distance envelope around ridges on the horizontal axis. The shaded area around the curve indicates 95\% confidence intervals for 10 runs per distance, demonstrating the low variation in coverage for comparatively small subsamples that enable fast runtimes. We expect a highly concave curve path to reflect a diminished increase in coverage with higher distances, as ridges should closely follow higher-density areas due to the way they are computed. The curve path in the figure clearly shows this behavior, with ridge envelopes covering 94\% of incidents at 0.1 miles for the whole city, quickly rising to 97.5\% and 98.5\% at 0.2 and 0.3 miles, respectively, and reaching 99\% coverage at about 0.6 miles.

In the lower right corner of the left-hand plot, we show a box-and-whiskers plot, with the upper and lower boundaries of the boxing indicating values within 1.5 times the interquartile range, the horizontal line intersecting the box denoting the median,and the off-standing `whiskers' indicating the minimum and maximum values \cite{McGill1978}. The number of iterations remains stable for different subsamples, demonstrating consistent convergence for subsampled sets in line with the narrow confidence shown in the primary plot. The right-hand plot of Fig.~\ref{fig:future_experiment} confirms the viability of density ridges as route templates, outperforming random routing within identified hot spot areas. As routes through a hot spot should offer more distance-based coverage by virtue of covering a larger area, one can reasonably expect center-only measurements to underperform both approaches as a sanity check, which the right-hand plot demonstrates.

\subsection{Mappability and route waypoints}

\begin{figure}
    \centering
    \includegraphics[width=\textwidth]{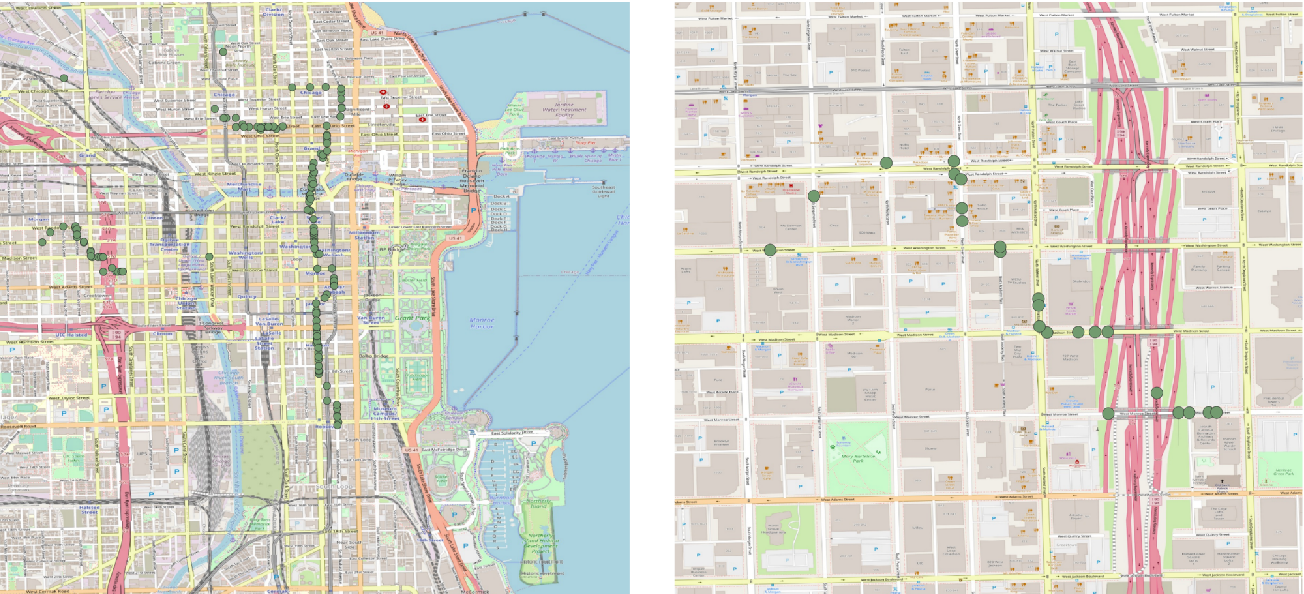}
    \caption{Route points plotted for parts of the top-5\% areas. Each green node is one highway location point registered on OpenStreetMap which is identified as being the closest to a ridge point. \textit{Left}: Large-scale view of the identified route points. \textit{Right:} Zoomed-in depiction of the left-most route shown in the left plot.}
    \label{fig:route_mapping}
\end{figure}

In order to allow for a translation of ridges to route guidelines, we make use of the R package `osmar' \cite{osmar2010}, which enables access to OpenStreetMap data. For each ridge point, we calculate the closest registered node on OpenStreetMap to allow us to make use of the underlying maps. From each of these neighbouring nodes, we then identify the nearest node which is located on a highway. These highway nodes can be seen in Fig.~\ref{fig:route_mapping} for the ridges in the top-5\% areas, as well as for a zoom-in of a single ridge segment. Due to OpenStreetMap being compatible with commonly used navigation systems, adding these maps into such systems is a straightforward approach. Patrols can then use these points to guide their routes while remaining flexible regarding their order or share of responsibility of area between the individual police officers.

\section{Discussion and limitations}
\label{sec:discussion}

In this exploratory study, we present DREDGE as a way to increase hot spot patrol efficiency and quality. Based on extensions from the field of cosmology, we make use of the SCMS algorithm for hot spot patrol routes. Our experiments show that optimized patrol templates cover about 94\% of incidents within 0.1 miles of ridges, reaching to about 99\% coverage at 0.6 miles. We implement multiple realizations of our experiments to investigate the stability of crime coverage with ridges based on past data. The corresponding results demonstrate relative stability within narrow confidence intervals across differing subsamples, validating the applicability of our approach for large-scale data.

Research on hot spots maintains that crime concentrates within a small geographic area \cite{weisburd2015law}. The widespread assumption when modeling hot spots is that the epicenter of the hot spots is where police attention should be focused. For example, the Pittsburgh Police Bureau used `putting cops on dots' for about 15 years \cite{gorr2015early}. Commonly employed density estimation methods can, however, obscure underlying features \cite{Eck2005}. The epicenter may be interpreted as a place to heavily patrol in lieu of surrounding areas that may deserve equal or more attention. Thus, this study's objective is to provide an effective alternative, and implies a refutation of previous assumptions about optimal patrol routes within hot spots to reduce crime through deterrent effects.

Empirical analyses that use KDE techniques or similar statistical modeling approaches often serve one function, namely guiding patrol routes. The issue of the epicenter misleading officers to focus patrols on the central area of a mode is a matter of identifying which places and routes will efficiently deliver deterrence effects. Therefore, in lieu of patrolling one stopping point that is criminogenic and identifiable, these ridges use the space around the problem places that lead to the epicenter, serving a dual function of patrolling the criminogenic locations and targeting high-risk locations across normal hot spots. Prior patrol-routing algorithm and selection work hone in on selecting strict routes that offer little route flexibility. Thus, in-between duties, this application is not meant to be the primary focus of patrols but rather an addition.

Our results show the coverage and potential efficiency of ridge patrolling. The ridges calculated with data from 2018 and 95\% confidence intervals in Fig.~\ref{fig:future_experiment} depict how ridge patrolling, hot spot patrols, and epicenter patrols work. They demonstrate the ridge patrolling method to be the most efficient, exposing nearly 95\% of Chicago's Part 1 crime incidents to directed patrols. If used solely in the epicenter of hot spots by thresholding the data to the highest 5\% of crime report densities, ridges still cover the majority of the crime area while providing patrol presence in the latter. Hence, using ridges is more equitable and responsive to the surrounding crime areas than regular epicenter patrolling. The selective or all-inclusive use of this method has the advantage and potential of being a high-efficiency crime prevention program. Furthermore, we argue that it can be a dynamic and widespread crime prevention measure across a city.

Our study is not without limitations. The methodology applied in this paper does not apply weights to problem places. Therefore, future work could focus on the application of spatial weights. In addition, this work assumes that the organization of patrol routes is implementable solely based on filament optimization, not considering community residents who may want to stop officers or demand more presence \cite{leigh2017predictive}. Given the fixed location of hot spots, the desires of residents, and the possible need to redraw routes due to calls for service or complex routes, such alterations should follow an as-close-as-possible alignment with ridges.

In addition, the program solely makes use of the geographic locations of prior incidents within a year frame. Thus, we do not account for the temporal dimension of the hot spots or incidents. However, an adaptation of this program by filtering and grouping incidents into time windows prior to running the program is feasible, which allows for more in-depth investigations. For follow-up research, we propose to combine such investigations with the separation of crime types to explore temporal changes in overall distributions, and weight shifts in types of incidents. Future work should, therefore, investigate such `hot times' and DREDGE for police work, offering yet another perspective to research on spatio-temporal crime patterns~\cite[see, for example,][]{ratcliffe2004hotspot, Newton2015, Malleson2015}.

Another limitation of this program is the lack of routing times for when officers should patrol each ridge, providing both spatial and temporal guidance. Building on the mention of weight shifts above, ridge segments for percentage-cut areas for different time windows could be weighted for their interest, allowing for time-dependent changes in route templates as well as a duration relying on expected crime density and types. On a more practical note, minimum durations for ridge segment traversal could be calculated through building a graph from points such as the ones covered in Section 4.3. By using either time estimates for average speed or, more advanced, linking the program into the route time prediction that navigation programs offer, traversal durations could be estimated for given segments.

While our implementation performs successful density ridge estimation in a matter of minutes, this requires subsampling from larger datasets of coordinates. As a rule of thumb, we recommend to use a minimum of 1,000 and a maximum of 10,000 data points to ensure representativeness and sufficiently fast runtimes. This is due to the algorithm's complexity being $\mathcal{O}(d \cdot |\theta|^2)$, meaning that it scales linearly with the number of dimensions, which is fixed to $|\theta| = 2$ in our case of latitude-longitude coordinates, but features a polynomial runtime due to the number of data points fed into the algorithm \cite{Ozertem2011}. While sample sizes are, in practice, influenced by both time constraints and the size of available datasets, details on effective sample sizes in geospatial dataset resampling can be found in the literature \cite{Griffith2005, Li2016}.

Bias in data is a general problem spanning many fields, which extends to geospatial coordinates. One prominent example is the phenomenon of over-policing and under-policing based on previous records, different socioeconomic status, and additional factors such as personal characteristics \cite{black1980manners}. Since our analysis is based on reported crime incidents, one important limitation of our work relates to previous research on disparities in crime reporting. This multi-faceted issue spans contextual factors in victim and offender characteristics influencing reporting \cite{Xie2012}, as well as localized reluctance of reporting crime incidents \cite{Slocum2010}. For this reason, practical implementations based on such data should always strive to take the risk of biases present in these datasets into account.

\section{Conclusion}
\label{sec:conclusion}

Optimizing police patrols, both city-wide and hot spots, remains an interest for researchers and practitioners alike. For this purpose, we show how recent advances in statistics and astronomy can be used to detect principal curves, or density ridges, in crime incident distributions to extract high-density paths. Current work focuses on the hot spot's epicenter, which decreases equitable patrol to surrounding areas and efficient hot spot patrols to the surrounding area. Overall, we provide a way to amend these issues through a density-based patrol optimization program.

Our study uses 2018 Part I crimes from the Chicago Police Department to demonstrate the patrol templates. Comparing this output to Part I crimes from early 2019, we observe that the majority of crime reports fall into narrow envelopes around identified structures. Thus, this program allocates resources around a hot spot, covering density regions. We argue that this allocates resources equitably and optimally to prevent crimes and reduce hot spots.The combination of hot spot mapping and DREDGE has the potential of being a high-efficiency crime prevention program providing a more responsive and optimal allocation of police resources than solely targeting epicenters of hot spots. We showcase the viability of our approach with intuitive visualizations, allowing for their combination with knowledge about city-specific traffic routes to plan effective patrols while remaining not overly constrained.

\section*{Acknowledgments}

We thank the City of Chicago and the Chicago Police Department for their open access data. We wish to express our gratitude to Nicholas Corsaro, Cory Haberman, and Monsuru Adepeju for helpful suggestions and comments. We also want to thank the two reviewers for their helpful comments in improving this paper.

\bibliographystyle{spbasic}
\bibliography{references.bib}

\begin{thebibliography}{82}
\providecommand{\natexlab}[1]{#1}
\providecommand{\url}[1]{{#1}}
\providecommand{\urlprefix}{URL }
\expandafter\ifx\csname urlstyle\endcsname\relax
  \providecommand{\doi}[1]{DOI~\discretionary{}{}{}#1}\else
  \providecommand{\doi}{DOI~\discretionary{}{}{}\begingroup
  \urlstyle{rm}\Url}\fi
\providecommand{\eprint}[2][]{\url{#2}}

\bibitem[{Al~Boni and Gerber(2016)}]{al2016automatic}
Al~Boni M, Gerber MS (2016) {Automatic optimization of localized kernel density
  estimation for hotspot policing}. In: 15th IEEE International Conference on
  Machine Learning and Applications, pp 32--38, \doi{10.1109/ICMLA.2016.0015}

\bibitem[{Barnett-Ryan et~al(2014)Barnett-Ryan, Langton, and
  Planty}]{barnett2014nation}
Barnett-Ryan C, Langton L, Planty M (2014) {The nation's two crime measures}.
  Tech. rep., Bureau of Justice Statistics \& Federal Bureau of Investigation,
  program report, U.S. Department of Justice, NCJ 246832

\bibitem[{{Bas} and {Erdogmus}(2011)}]{Bas2011}
{Bas} E, {Erdogmus} D (2011) {Principal curves as skeletons of tubular
  objects}. Neuroinformatics 9(2):181--191, \doi{10.1007/s12021-011-9105-2}

\bibitem[{Black(1980)}]{black1980manners}
Black D (1980) {The manners and customs of the police}. New York: Academic
  Press

\bibitem[{Bodily(1978)}]{bodily1978police}
Bodily SE (1978) Police sector design incorporating preferences of interest
  groups for equality and efficiency. J Manag Sci 24(12):1301--1313,
  \doi{10.1287/mnsc.24.12.1301}

\bibitem[{Bowers et~al(2004)Bowers, Johnson, and Pease}]{Bowers2004}
Bowers KJ, Johnson SD, Pease K (2004) Prospective hot-spotting: The future of
  crime mapping? Br J Criminol 44(5):641--658, \doi{10.1093/bjc/azh036}

\bibitem[{Braga et~al(2012)Braga, Papachristos, and Hureau}]{braga2012hot}
Braga A, Papachristos A, Hureau D (2012) {Hot spots policing effects on crime}.
  Campbell Syst Rev 8(8):1--96, \doi{10.4073/csr.2012.8}

\bibitem[{Braga(2007)}]{braga2007policing}
Braga AA (2007) {Policing crime hot spots}. In: Preventing Crime, New York:
  Springer Publishing, pp 179--192, \doi{10.1007/1-4020-4244-2_12}

\bibitem[{Braga et~al(2010)Braga, Papachristos, and
  Hureau}]{braga2010concentration}
Braga AA, Papachristos AV, Hureau DM (2010) {The concentration and stability of
  gun violence at micro places in {B}oston, 1980-2008}. J Quant Criminol
  26(1):33--53, \doi{10.1007/s10940-009-9082-x}

\bibitem[{Braga et~al(2014)Braga, Papachristos, and Hureau}]{braga2014effects}
Braga AA, Papachristos AV, Hureau DM (2014) {The effects of hot spots policing
  on crime: An updated systematic review and meta-analysis}. Justice Q
  31(4):633--663, \doi{10.1080/07418825.2012.673632}

\bibitem[{{Camacho-Collados} and {Liberatore}(2015)}]{Camacho-Collados2015}
{Camacho-Collados} M, {Liberatore} F (2015) {A decision support system for
  predictive police patrolling}. Decis Support Syst 75:25--37,
  \doi{10.1016/j.dss.2015.04.012}

\bibitem[{Caplan et~al(2011)Caplan, Kennedy, and Miller}]{caplan2011risk}
Caplan JM, Kennedy LW, Miller J (2011) Risk terrain modeling: Brokering
  criminological theory and {GIS} methods for crime forecasting. Justice Q
  28(2):360--381, \doi{10.1080/07418825.2010.486037}

\bibitem[{Chainey et~al(2008{\natexlab{a}})Chainey, Tompson, and
  Uhlig}]{chainey2008utility}
Chainey S, Tompson L, Uhlig S (2008{\natexlab{a}}) {The utility of hotspot
  mapping for predicting spatial patterns of crime}. Secur J 21(1-2):4--28,
  \doi{10.1057/palgrave.sj.8350066}

\bibitem[{Chainey et~al(2008{\natexlab{b}})Chainey, Tompson, and
  Uhlig}]{Chainey2008}
Chainey S, Tompson L, Uhlig S (2008{\natexlab{b}}) The utility of hotspot
  mapping for predicting spatial patterns of crime. Secur J 21(1):4--28,
  \doi{10.1057/palgrave.sj.8350066}

\bibitem[{Chawathe(2007)}]{chawathe2007organizing}
Chawathe SS (2007) {Organizing hot-spot police patrol routes}. In: 2007 IEEE
  International Conference on Intelligence and Security Informatics, pp 79--86,
  \doi{10.1109/ISI.2007.379538}

\bibitem[{Chen et~al(2015)Chen, Cheng, and Wise}]{chen2015designing}
Chen H, Cheng T, Wise S (2015) Designing daily patrol routes for policing based
  on {ANT} colony algorithm. ISPRS Ann Photogrammetry, Remote Sens Spat Inf Sci
  2:103--109, \doi{10.5194/isprsannals-II-4-W2-103-2015}

\bibitem[{Chen et~al(2017)Chen, Cheng, and Wise}]{chen2017developing}
Chen H, Cheng T, Wise S (2017) Developing an online cooperative police patrol
  routing strategy. Comp Env Urban Sys 62:19--29,
  \doi{10.1016/j.compenvurbsys.2016.10.013}

\bibitem[{{Chen} et~al(2015{\natexlab{a}}){Chen}, {Genovese}, and
  {Wasserman}}]{Chen2015c}
{Chen} YC, {Genovese} CR, {Wasserman} L (2015{\natexlab{a}}) {Asymptotic theory
  for density ridges}. Ann Stat 43(5):1896--1928

\bibitem[{{Chen} et~al(2015{\natexlab{b}}){Chen}, {Ho}, {Freeman}, {Genovese},
  and {Wasserman}}]{Chen2015a}
{Chen} YC, {Ho} S, {Freeman} PE, {Genovese} CR, {Wasserman} L
  (2015{\natexlab{b}}) {Cosmic web reconstruction through density ridges:
  Method and algorithm}. Mon Notices Royal Astron Soc 454:1140--1156,
  \doi{10.1093/mnras/stv1996}

\bibitem[{{Chen} et~al(2015{\natexlab{c}}){Chen}, {Ho}, {Tenneti},
  {Mandelbaum}, {Croft}, {DiMatteo}, {Freeman}, {Genovese}, and
  {Wasserman}}]{Chen2015b}
{Chen} YC, {Ho} S, {Tenneti} A, {Mandelbaum} R, {Croft} R, {DiMatteo} T,
  {Freeman} PE, {Genovese} CR, {Wasserman} L (2015{\natexlab{c}})
  {Investigating galaxy-filament alignments in hydrodynamic simulations using
  density ridges}. Mon Notices Royal Astron Soc 454:3341--3350,
  \doi{10.1093/mnras/stv2260}

\bibitem[{{Chen} et~al(2016){Chen}, {Ho}, {Brinkmann}, {Freeman}, and
  {Wasserman}}]{Chen2016}
{Chen} YC, {Ho} S, {Brinkmann} J, {Freeman} PEP, {Wasserman} L (2016) {Cosmic
  web reconstruction through density ridges: Catalogue}. Mon Notices Royal
  Astron Soc 461:3896--3909, \doi{10.1093/mnras/stw1554}

\bibitem[{{Chen} et~al(2017){Chen}, {Ho}, {Mandelbaum}, {Bahcall},
  {Brownstein}, {Freeman}, {Genovese}, {Schneider}, and {Wasserman}}]{Chen2017}
{Chen} YC, {Ho} S, {Mandelbaum} R, {Bahcall} NA, {Brownstein} JR, {Freeman} PE,
  {Genovese} CR, {Schneider} DP, {Wasserman} L (2017) {Detecting effects of
  filaments on galaxy properties in the {S}loan {D}igital {S}ky {S}urvey
  {III}}. Mon Notices Royal Astron Soc 466:1880--1893,
  \doi{10.1093/mnras/stw3127}

\bibitem[{Chevaleyre(2004)}]{chevaleyre2004theoretical2}
Chevaleyre Y (2004) {Theoretical analysis of the multi-agent patrolling
  problem}. In: 2004 IEEE/WIC/ACM International Conference on Intelligent Agent
  Technology, pp 302--308, \doi{10.1109/IAT.2004.1342959}

\bibitem[{Corsaro et~al(2019)Corsaro, Engel, Herold, and
  Yildirim}]{corsaro2019implementing}
Corsaro N, Engel RS, Herold TD, Yildirim M (2019) Implementing gang \& gun
  violence reduction strategies in las vegas, nevada: Hot spots evaluation
  results

\bibitem[{{Cover} and {Hart}(1967)}]{Cover1967}
{Cover} TM, {Hart} PE (1967) {Nearest neighbor pattern classification}. IEEE
  Trans Inf Theory 13(1):21--27, \doi{10.1109/TIT.1967.1053964}

\bibitem[{Eck(1997)}]{eck1997those}
Eck JE (1997) What do those dots mean? mapping theories with data. In: Crime
  mapping and crime prevention, New York: Criminal Justice Press, pp 379--406

\bibitem[{Eck and Guerette(2012)}]{eck2012place}
Eck JE, Guerette RT (2012) Place-based crime prevention: Theory, evidence, and
  policy. The Oxford handbook of crime prevention pp 354--383,
  \doi{10.1093/oxfordhb/9780195398823.013.0018}

\bibitem[{{Eck} et~al(2005){Eck}, {Chainey}, {Cameron}, {Leitner}, and
  {Wilson}}]{Eck2005}
{Eck} JE, {Chainey} S, {Cameron} JG, {Leitner} M, {Wilson} RE (2005) {Mapping
  crime: Understanding hot spots}, 1st edn. Washington, D.C.: Office of Justice
  Programs, National Institute of Justice

\bibitem[{Eugster and Schlesinger(2013)}]{osmar2010}
Eugster MJA, Schlesinger T (2013) {osmar: OpenStreetMap and R}. {The R Journal}
  5(1):53--63, \doi{10.32614/RJ-2013-005}

\bibitem[{{Fukunaga} and {Hostetler}(1975)}]{Fukunaga1975}
{Fukunaga} K, {Hostetler} LD (1975) {The estimation of the gradient of a
  density function, with applications in pattern recognition}. IEEE Trans Inf
  Theory 21(1):32--40, \doi{10.1109/TIT.1975.1055330}

\bibitem[{Furtado et~al(2006)Furtado, Melo, Menezes, and
  Belchior}]{furtado2006using}
Furtado V, Melo A, Menezes R, Belchior M (2006) {Using self-organization in an
  agent framework to model criminal activity in response to police patrol
  routes}. In: 2006 Florida Artificial Intelligence Research Society
  Conference, pp 68--73

\bibitem[{{Furtado} et~al(2009){Furtado}, {Melo}, {Coelho}, {Menezes}, and
  {Perrone}}]{Furtado2009}
{Furtado} V, {Melo} A, {Coelho} ALV, {Menezes} R, {Perrone} R (2009) {A
  bio-inspired crime simulation model}. Decis Support Syst 48(1):282--292,
  \doi{10.1016/j.dss.2009.08.008}

\bibitem[{{Genovese} et~al(2012){Genovese}, {Perone-Pacifico}, {Verdinelli},
  and {Wasserman}}]{Genovese2012}
{Genovese} CR, {Perone-Pacifico} M, {Verdinelli} I, {Wasserman} L (2012) {The
  geometry of nonparametric filament estimation}. J Am Stat Assoc
  107(498):788--799, \doi{10.1080/01621459.2012.682527}

\bibitem[{{Genovese} et~al(2014){Genovese}, {Perone-Pacifico}, {Verdinelli},
  and {Wasserman}}]{Genovese2014}
{Genovese} CR, {Perone-Pacifico} M, {Verdinelli} I, {Wasserman} L (2014)
  {Nonparametric ridge estimation}. Ann Statist 42(4):1511--1545,
  \doi{10.1214/14-AOS1218}

\bibitem[{{Ghassabeh} et~al(2013){Ghassabeh}, Linder, and
  Takahara}]{Ghassabeh2013}
{Ghassabeh} YA, Linder T, Takahara G (2013) {On some convergence properties of
  the subspace constrained mean shift}. Pattern Recognit 46(11):3140--3147,
  \doi{10.1016/j.patcog.2013.04.014}

\bibitem[{Gorr and Lee(2015)}]{gorr2015early}
Gorr WL, Lee Y (2015) Early warning system for temporary crime hot spots.
  Journal of Quantitative Criminology 31(1):25--47,
  \doi{10.1007/s10940-014-9223-8}

\bibitem[{Griffith(2005)}]{Griffith2005}
Griffith DA (2005) {Effective geographic sample size in the presence of spatial
  autocorrelation}. Ann Am Assoc Geogr 95(4):740--760,
  \doi{10.1111/j.1467-8306.2005.00484.x}

\bibitem[{Haberman(2017)}]{haberman2017overlapping}
Haberman CP (2017) {Overlapping hot spots? Examination of the spatial
  heterogeneity of hot spots of different crime types}. Criminol Public Policy
  16(2):633--660, \doi{10.1111/1745-9133.12303}

\bibitem[{{He} et~al(2017){He}, {Alam}, {Ferraro}, {Chen}, and {Ho}}]{He2017}
{He} S, {Alam} S, {Ferraro} S, {Chen} YC, {Ho} S (2017) {The detection of the
  imprint of filaments on cosmic microwave background lensing}. Nat Astron
  2(5):401--406, \doi{10.1038/s41550-018-0426-z}

\bibitem[{Hendel et~al(2019)Hendel, Johnston, Patra, and Sen}]{Hendel2019}
Hendel D, Johnston KV, Patra RK, Sen B (2019) {A machine-vision method for
  automatic classification of stellar halo substructure}. Mon Notices Royal
  Astron Soc 486(3):3604--3616, \doi{10.1093/mnras/stz1107}

\bibitem[{{Inman}(1835)}]{Inman1835}
{Inman} JW (1835) {Navigation and nautical astronomy for the use of {B}ritish
  seamen}, 3rd edn. London: W. Woodward, C. \& J. Rivington

\bibitem[{Koper(1995)}]{koper1995just}
Koper CS (1995) {Just enough police presence: Reducing crime and disorderly
  behavior by optimizing patrol time in crime hot spots}. Justice Q
  12(4):649--672, \doi{10.1080/07418829500096231}

\bibitem[{Kringen et~al(2017)Kringen, Sedelmaier, and
  Elink-Schuurman-Laura}]{kringen2017assessing}
Kringen JA, Sedelmaier CM, Elink-Schuurman-Laura KD (2017) Assessing the
  relevance of statistics and crime analysis courses for working crime
  analysts. J Criminal Justice Education 28(2):155--173,
  \doi{10.1080/10511253.2016.1192211}

\bibitem[{Leigh et~al(2017)Leigh, Dunnett, and Jackson}]{leigh2017predictive}
Leigh J, Dunnett S, Jackson L (2017) {Predictive police patrolling to target
  hotspots and cover response demand}. Ann Oper Res pp 1--16,
  \doi{10.1007/s10479-017-2528-x}

\bibitem[{Li et~al(2016)Li, Griffith, and Becker}]{Li2016}
Li B, Griffith DA, Becker B (2016) {Spatially simplified scatterplots for large
  raster datasets}. Geo Spat Inf Sci 19(2):81--93,
  \doi{10.1080/10095020.2016.1179441}

\bibitem[{Li et~al(2011)Li, Jiang, Duan, Dong, Hu, and Sun}]{li2011police}
Li L, Jiang Z, Duan N, Dong W, Hu K, Sun W (2011) {Police patrol service
  optimization based on the spatial pattern of hotspots}. In: 2011 IEEE
  International Conference on Service Operations, Logistics and Informatics, pp
  45--50, \doi{10.1109/SOLI.2011.5986526}

\bibitem[{Liberatore et~al(2020)Liberatore, Camacho-Collados, and
  Vitoriano}]{liberatore2020police}
Liberatore F, Camacho-Collados M, Vitoriano B (2020) Police districting
  problem: Literature review and annotated bibliography. In: Optimal
  Districting and Territory Design, New York: Springer Publishing, pp 9--29

\bibitem[{Linning and Eck(2018)}]{linning2018weak}
Linning SJ, Eck JE (2018) Weak intervention backfire and criminal hormesis: Why
  some otherwise effective crime prevention interventions can fail at low
  doses. The British Journal of Criminology 58(2):309--331,
  \doi{10.1093/bjc/azx019}

\bibitem[{Malleson and Andresen(1978)}]{Malleson2015}
Malleson N, Andresen MA (1978) Spatio-temporal crime hotspots and the ambient
  population. Crime Sci 4:10, \doi{10.1186/s40163-015-0025-6}

\bibitem[{Mamalian et~al(1999)Mamalian, La~Vigne et~al}]{mamalian1999use}
Mamalian CA, La~Vigne NG, et~al (1999) {The use of computerized crime mapping
  by law enforcement: Survey results}. Washington, D.C.: U.S. Dept. of Justice,
  Office of Justice Programs, National Institute of Justice

\bibitem[{Marchant et~al(2018)Marchant, Lu, and Cripps}]{marchant2018cox}
Marchant R, Lu D, Cripps S (2018) {Cox {B}ayesian optimization for police
  patrolling}. In: 32nd Annual Conference on Neural Information Processing
  Systems

\bibitem[{Mastrofski and Fridell(2015)}]{mastrofski2015police}
Mastrofski SD, Fridell L (2015) Police departments’ adoption of innovative
  practices

\bibitem[{McGill et~al(1978)McGill, Tukey, and Larsen}]{McGill1978}
McGill R, Tukey JW, Larsen WA (1978) {Variations of box plots}. Am Stat
  32(1):12--16

\bibitem[{Melo et~al(2005)Melo, Belchior, and Furtado}]{melo2005analyzing}
Melo A, Belchior M, Furtado V (2005) {Analyzing police patrol routes by
  simulating the physical reorganization of agents}. In: International Workshop
  on Multi-Agent Systems and Agent-Based Simulation, pp 99--114,
  \doi{10.1007/11734680_8}

\bibitem[{Menton(2008)}]{menton2008bicycle}
Menton C (2008) {Bicycle patrols: An underutilized resource}. Policing
  31(1):93--108, \doi{10.1108/13639510810852594}

\bibitem[{{Miao} et~al(2014){Miao}, {Wang}, {Shi}, and {Wu}}]{Miao2014}
{Miao} Z, {Wang} B, {Shi} W, {Wu} H (2014) {A method for accurate road
  centerline extraction from a classified image}. IEEE J Sel Top Appl Earth Obs
  Remote Sens 7(12):4762--4771, \doi{10.1109/JSTARS.2014.2309613}

\bibitem[{Mitchell(1972)}]{mitchell1972optimal}
Mitchell PS (1972) Optimal selection of police patrol beats. J Crim L
  Criminology \& Police Sci 63:577, \doi{10.2307/1141814}

\bibitem[{{Moews} et~al(2020){Moews}, {Schmitz}, {Lawler}, {Zuntz}, {Malz}, {de
  Souza}, {Vilalta}, {Krone-Martins}, and {Ishida}}]{Moews2020}
{Moews} B, {Schmitz} MA, {Lawler} AJ, {Zuntz} J, {Malz} AI, {de Souza} RS,
  {Vilalta} R, {Krone-Martins} A, {Ishida} EEO (2020) {Ridges in the {D}ark
  {E}nergy {S}urvey for cosmic trough identification}. arXiv e-prints
  arXiv:2005.08583

\bibitem[{Newton and Felson(1978)}]{Newton2015}
Newton A, Felson M (1978) Editorial: crime patterns in time and space: the
  dynamics of crime opportunities in urban areas. Crime Sci 4:11,
  \doi{10.1186/s40163-015-0025-6}

\bibitem[{{Ozertem} and {Erdogmus}(2011)}]{Ozertem2011}
{Ozertem} U, {Erdogmus} D (2011) {Locally defined principal curves and
  surfaces}. J Mach Learn Res 12:1249--1286

\bibitem[{Paruchuri et~al(2008)Paruchuri, Pearce, Marecki, Tambe, Ordonez, and
  Kraus}]{paruchuri2008playing}
Paruchuri P, Pearce JP, Marecki J, Tambe M, Ordonez F, Kraus S (2008) {Playing
  games for security: An efficient exact algorithm for solving {B}ayesian
  {S}tackelberg games}. In: 7th International Joint Conference on Autonomous
  Agents and Multiagent Systems, Vol. 2, pp 895--902

\bibitem[{{Parzen}(1962)}]{Parzen1962}
{Parzen} E (1962) {On estimation of a probability density function and mode}.
  Ann Math Statist 33(3):1065--1076, \doi{10.1214/aoms/1177704472}

\bibitem[{Piyadasun et~al(2017)Piyadasun, Kalansuriya, Gangananda, Malshan,
  Bandara, and Marru}]{piyadasun2017rationalizing}
Piyadasun T, Kalansuriya B, Gangananda M, Malshan M, Bandara HD, Marru S (2017)
  Rationalizing police patrol beats using heuristic-based clustering. In: 2017
  Moratuwa Engineering Research Conference (MERCon), IEEE, pp 431--436,
  \doi{MERCon.2017.7980523}

\bibitem[{{Qiao} and {Polonik}(2016)}]{Qiao2016}
{Qiao} W, {Polonik} W (2016) {Theoretical analysis of nonparametric filament
  estimation}. Ann Statist 44(3):1269--1297, \doi{10.1214/15-AOS1405}

\bibitem[{{Ratcliffe}(2010)}]{ratcliffe2010crime}
{Ratcliffe} J (2010) {Crime mapping: Spatial and temporal challenges}. In:
  Handbook of quantitative criminology, New York: Springer Publishing, pp
  5--24, \doi{10.1007/978-0-387-77650-7_2}

\bibitem[{Ratcliffe(2004{\natexlab{a}})}]{ratcliffe2004crime}
Ratcliffe JH (2004{\natexlab{a}}) {Crime mapping and the training needs of law
  enforcement}. Eur J Crim Policy Res 10(1):65--83,
  \doi{10.1023/B:CRIM.0000037550.40559.1c}

\bibitem[{Ratcliffe(2004{\natexlab{b}})}]{ratcliffe2004hotspot}
Ratcliffe JH (2004{\natexlab{b}}) The hotspot matrix: A framework for the
  spatio-temporal targeting of crime reduction. Police Pract Res 5(1):5--23,
  \doi{10.1080/1561426042000191305}

\bibitem[{Reis et~al(2006)Reis, Melo, Coelho, and Furtado}]{reis2006gapatrol}
Reis D, Melo A, Coelho AL, Furtado V (2006) {{GAP}atrol: An evolutionary
  multiagent approach for the automatic definition of hotspots and patrol
  routes}. In: Advances in Artificial Intelligence - IBERAMIA-SBIA 2006, New
  York: Springer Publishing, pp 118--127, \doi{10.1007/11874850_16}

\bibitem[{{Rosenblatt}(1956)}]{Rosenblatt1956}
{Rosenblatt} M (1956) {Remarks on some nonparametric estimates of a density
  function}. Ann Math Statist 27(3):832--837, \doi{10.1214/aoms/1177728190}

\bibitem[{Sherman and Weisburd(1995)}]{sherman1995general}
Sherman LW, Weisburd D (1995) {General deterrent effects of police patrol in
  crime ``hot spots'': A randomized, controlled trial}. Justice Q
  12(4):625--648, \doi{10.1080/07418829500096221}

\bibitem[{Sherman et~al(1989)Sherman, Gartin, and Buerger}]{sherman1989hot}
Sherman LW, Gartin PR, Buerger ME (1989) {Hot spots of predatory crime: Routine
  activities and the criminology of place}. Criminology 27(1):27--56,
  \doi{10.1111/j.1745-9125.1989.tb00862.x}

\bibitem[{{Slocum} et~al(2010){Slocum}, {Taylor}, {Brick}, and
  {Esbensen}}]{Slocum2010}
{Slocum} LA, {Taylor} TJ, {Brick} BT, {Esbensen} FA (2010) {Neighborhood
  structural characteristics, individual-level attitudes, and youths' crime
  reporting intentions}. Criminology 48(4):1063--1100,
  \doi{10.1111/j.1745-9125.2010.00212.x}

\bibitem[{Telep et~al(2014)Telep, Mitchell, and Weisburd}]{telep2014much}
Telep CW, Mitchell RJ, Weisburd D (2014) {How much time should the police spend
  at crime hot spots? Answers from a police agency directed randomized field
  trial in {S}acramento, {C}alifornia}. Justice Q 31(5):905--933,
  \doi{10.1080/07418825.2012.710645}

\bibitem[{{Wasserman}(2018)}]{Wassermann2018}
{Wasserman} L (2018) {Topological data analysis}. Annu Rev Stat Appl
  5(1):501--532, \doi{10.1146/annurev-statistics-031017-100045}

\bibitem[{Weisburd(2015)}]{weisburd2015law}
Weisburd D (2015) {The law of crime concentration and the criminology of
  place}. Criminology 53(2):133--157, \doi{10.1111/1745-9125.12070}

\bibitem[{Weisburd and Lum(2005)}]{weisburd2005diffusion}
Weisburd D, Lum C (2005) {The diffusion of computerized crime mapping in
  policing: Linking research and practice}. Police Pract Res 6(5):419--434,
  \doi{10.1080/15614260500433004}

\bibitem[{Weisburd and Majmundar(2018)}]{national2018proactive}
Weisburd D, Majmundar MK (2018) Proactive policing: Effects on crime and
  communities. Committee on proactive policing: Effects on crime, communities,
  and civil liberties. Washington, D.C.: National Academies Press

\bibitem[{Weisburd et~al(2004)Weisburd, Bushway, Lum, and
  Yang}]{weisburd2004trajectories}
Weisburd D, Bushway S, Lum C, Yang SM (2004) {Trajectories of crime at places:
  A longitudinal study of street segments in the city of {S}eattle}.
  Criminology 42(2):283--322, \doi{10.1111/j.1745-9125.2004.tb00521.x}

\bibitem[{Williams and Coupe(2017)}]{williams2017frequency}
Williams S, Coupe T (2017) Frequency vs. length of hot spots patrols: a
  randomised controlled trial. Cambridge Journal of Evidence-Based Policing
  1(1):5--21, \doi{10.1007/s41887-017-0003-1}

\bibitem[{{Williamson} et~al(1999){Williamson}, {McLafferty}, {Goldsmith},
  {Mallenkopf}, and {McGuire}}]{Williamson1999}
{Williamson} D, {McLafferty} S, {Goldsmith} V, {Mallenkopf} J, {McGuire} P
  (1999) {A better method to smooth crime incident data}. ESRI ArcUser Magazine
  January--March 1999:1--5

\bibitem[{{Xie} and {Lauritsen}(2012)}]{Xie2012}
{Xie} M, {Lauritsen} JL (2012) {Racial context and crime reporting: A test of
  {B}lack's stratification hypothesis}. J Quant Criminol 28(2):265--293,
  \doi{10.1007/s10940-011-9140-z}

\bibitem[{{Xue} and {Brown}(2006)}]{Xue2006}
{Xue} Y, {Brown} DE (2006) {Spatial analysis with preference specification of
  latent decision makers for criminal event prediction}. Decis Support Syst
  41(3):560--573, \doi{10.1016/j.dss.2004.06.007}

\end{thebibliography}

\end{document}